\documentclass[prb,twocolumn,showpacs,preprintnumbers,amsmath,amssymb,superscriptaddress]{revtex4-1}

\usepackage{graphicx}
\usepackage{dcolumn}
\usepackage{bm}
\usepackage{mathbbol}
\usepackage{color}
\usepackage[FIGTOPCAP,raggedright,nooneline]{subfigure}

\newcommand{\bra}[1]{\langle #1 |}
\newcommand{\ket}[1]{| #1 \rangle}

\newcommand{\Bra}{\langle}
\newcommand{\Ket}{\rangle}

\newcommand{\ignore}[1]{}

\begin{document}


\title{Self-Consistent Projection Operator Theory for Quantum Many-Body Systems}

\author{Peter Degenfeld-Schonburg}
\email{peter.degenfeld-schonburg@ph.tum.de}
\affiliation{Technische Universit{\"a}t M{\"u}nchen, Physik Department, James Franck Str., 85748 Garching, Germany}
\author{Michael J. Hartmann}
\email{m.j.hartmann@hw.ac.uk}
\affiliation{Institute of Photonics and Quantum Sciences, Heriot-Watt University, Edinburgh, EH14 4AS, United Kingdom}
\affiliation{Technische Universit{\"a}t M{\"u}nchen, Physik Department, James Franck Str., 85748 Garching, Germany}

\date{\today}

\begin{abstract}
We derive an exact equation of motion for the reduced density matrices of individual subsystems of quantum many-body systems of any 
lattice dimension and arbitrary system size. Our projection operator based theory yields a highly efficient analytical and numerical approach. 
Besides its practical use it provides a novel interpretation and systematic extension of mean-field approaches and an adaption of open quantum systems 
theory to settings where a dynamically evolving environment has to be taken into account. We show its high accuracy for two significant classes of complex 
quantum many-body dynamics, unitary evolutions of non-equilibrium states in closed and stationary states in driven-dissipative systems.
\end{abstract}

\maketitle

%
%
\section{Introduction}
Quantum many-body systems give rise to a number of 
intriguing phenomena such as quantum phase transitions \cite{Sachdev11},
topological insulators \cite{Hasan10,Nayak08} 
or high-temperature superconductivity \cite{Leggett06}.
Yet, their description is a formidable challenge as the dimension of the Hilbert space grows exponentially with number of constituents. 
The huge number of degrees of freedom thus renders an exact description in general infeasible, even if one resorts to numerical approaches.
Exceptions to this intractability are quantum systems that do not explore their entire Hilbert space, where numerical optimization approaches such as 
the Density Matrix Renormalization Group \cite{Schollwoeck05} become efficient descriptions.
Alternatively one may aim for only obtaining the information of interest about the quantum state of 
the entire system and try to find accurate and efficient approximations for the sought quantities.
Mean-field approaches \cite{Kadanoff09} can be understood as representatives of this strategy as they only predict properties of 
a single constituent of the many-body system \cite{Sachdev11,Fisher89}.

Equations of motion for the part of the quantum state that is of interest to the researcher have been derived in the context of open quantum systems 
where the density matrix of the entire system is split into a 'relevant' part describing the system and a complementary 'irrelevant' part with the help of
the Mori projector \cite{Mori65,Zwanzig01,Breuer07}.

Here, we introduce an approach to the calculation of local properties
of a quantum many-body system by defining a time dependent projection operator that may be viewed as a generalization of the Mori projector \cite{Mori65}. 
Based on this projector, which we coin self-consistent Mori projector (c-MoP), we are able to derive an integro-differential 
equation that shares similarities with a Nakajima-Zwanzig equation  \cite{Zwanzig01,Breuer07} and exactly describes the 
dynamics of the reduced density matrix of one subsystem (or a cluster of subsystems) of a quantum many-body system.

Our theory is thus capable of describing stationary states and dynamical evolutions for any situation in which one is only interested in the physics of a part of the system under study.
We thus expect applications of our theory to be very useful for quantum few- and many-body systems.
Most notably, it efficiently predicts non-equilibrium dynamics for very long times,
applies to two- and higher-dimensional systems in the same way as to one-dimensional ones, and can directly and efficiently calculate stationary states of many-body systems with dissipation.
Besides these applications our technique generalizes the theory of open quantum systems to scenarios beyond the Markov or thermal equilibrium regimes as it takes the back action onto the environment 
into account.

An increasing number of experimental settings, including arrays of 
Josephson junctions \cite{FZ01}, ultra-cold atoms
\cite{BDZ07}, ion traps \cite{ion,Barreiro11} and arrays 
of coupled cavities \cite{Hartmann06}, offer the possibility
to generate effective many-particle systems and hence trigger substantial research activity.

One prominent application of these systems are investigations of the unitary dynamics of non-equilibrium states in closed systems \cite{Kollath07,Trotzky12}. 
As a first test of the performance of our method we thus apply it to calculate local properties of time-evolving non-equilibrium states in closed systems.
We find that it predicts these quantities with very high accuracy for a time range that strongly increases with the size of the considered subsystem and for small 
subsystems already becomes comparable to the time range reached with the time dependent Density Matrix Renormalization Group (t-DMRG).

In many experimental situations, the samples will however suffer from decoherence and dissipation. 
Hence dissipative and driven-dissipative quantum many-body systems are currently receiving enormous interest 
in the search for strongly correlated steady states and non-equilibrium analogs of quantum phase transitions \cite{Barreiro11,Prosen08,Hartmann10,Nissen12,Diehl2008,Diehl2010,Diehl11,Rivas12}.
As a second test we thus apply our method to driven and dissipative quantum many-body systems and find that 
it predicts the values of local quantities with very good accuracy.

The remainder of the paper is organized as follows.
In section \ref{Sec:theory}, we introduce the self-consistent Mori projector (c-MoP) and derive an exact equation of motion for the reduced density matrices 
of individual subsystems of a quantum many-body system.
In section \ref{Sec:tests} we show the applicability and accuracy of our method for both, the unitary dynamics in closed quantum systems \ref{Sec:tests closed}, and 
non-unitary dynamics in one-dimensional \ref{Sec:tests open} and two-dimensional \ref{Sec:tests 2d} open quantum systems. 
Finally we give our conclusions and an outlook in section \ref{conclusions}.

\section{Self-consistent projection operator theory}\label{Sec:theory}

\begin{figure}
\centering
\includegraphics[width=\columnwidth]{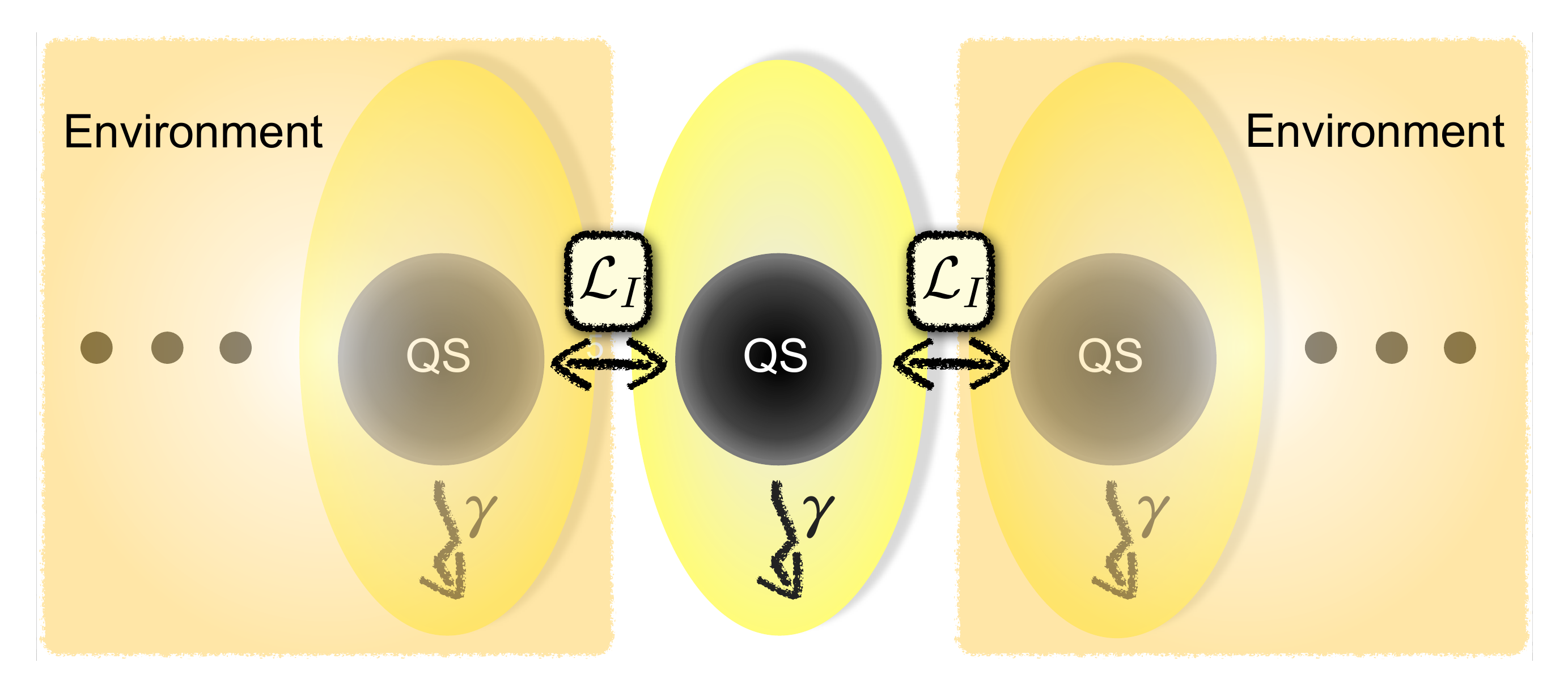}
\caption{\label{notion of a QMS} Illustration of our approach for a one-dimensional lattice. 
We consider a quantum many-body system where each subsystem (QS) has some unitary and potentially some 
non-unitary dynamics (indicated by the rate $\gamma$). The subsystems are coupled via the interaction $\mathcal L_I$. Within our theory we pick one QS of interest and 
trace out the remaining constituents.}
\end{figure}

We consider a quantum many-body system of $N$-partite structure for which the density matrix $R(t)$ of the entire system obeys the Liouville equation of motion, 
\begin{equation}\label{EOM main text}
\dot R(t)=\mathcal{L}R(t),
\end{equation}
where the dot denotes a time derivative and $\mathcal{L}=\mathcal{L}_{0}+\mathcal{L}_{I}$ with $\mathcal{L}_{0} = \sum_{n=1}^N\mathcal{L}_{n}$.
Here, the superoperators $\mathcal{L}_{n}$ describe the free dynamics 
of the $n$-th constituent and $\mathcal{L}_{I}$ accounts for the interaction between any of the $N$ parts. 
Importantly, $\mathcal{L}$ may feature non-unitary terms \cite{Lindblad}.
We are only interested in the properties of one subsystem, say subsystem $n_{0}$, and discard information about the remaining subsystems, see Fig.~\ref{notion of a QMS}
for an illustration. 
For these aims it suffices to determine the reduced density matrix
$\rho_{n_{0}}(t) = \text{Tr}_{\not{n_{0}}} R(t)$, where $\text{Tr}_{\not{n}}$ denotes the trace over all $N$ constituents but the $n$-th. 
To derive an equation of motion for $\rho_{n_{0}}(t)$ we define the time dependent projection operator
\begin{equation}\label{eq:MoriP-main-text}
  P_{t}^{n_{0}} ( \cdot ) =\rho_{\not{n_{0}}}(t)\otimes \text{Tr}_{\not{n_{0}}}(\cdot),
\end{equation}
where $\rho_{\not{n_{0}}}(t)$ is a reference density matrix for the degrees of 
freedom that have been traced out. $P_t^{n_{0}} R(t) = \rho_{\not{n_{0}}}(t)\otimes \rho_{n_{0}}(t)$ yields our quantity of interest $\rho_{n_{0}}(t)$.
In the language of open quantum systems \cite{Breuer07} one would call the subsystem $n_{0}$ the ``system'' and the remaining 
subsystems, $n \neq n_{0}$ the ``environment'' and for cases where $\rho_{\not{n_{0}}}$ is constant, the projector $P_{t}^{n_{0}}$ would reduce to the celebrated Mori projector \cite{Mori65}.
As we aim at deriving an equation of motion for the reduced density matrices $\rho_{n}(t) = \text{Tr}_{\not{n}} R(t)$ only,
we take $\rho_{\not{n_{0}}}(t)$ to read,
\begin{equation}\label{dens matr env main text}
\rho_{\not{n_{0}}}(t)=\bigotimes_{n\not{=} n_{0}}\rho_n(t) \;\;\text{with}\;\;\rho_n(t)=\text{Tr}_{\not{n}} R(t).
\end{equation}
%
In contrast to standard approaches \cite{Breuer07}, the reference density matrix 
$\rho_{\not{n_{0}}}(t)$, is
in our approach determined consistently from the state $R(t)$ of the entire setup.
This general approach requires to allow for a time dependence of the projector $P_{t}^{n_{0}}$.
A more general choice for $\rho_{\not{n_{0}}}$ could be $\rho_{\not{n_{0}}} = \text{Tr}_{n_{0}} R(t)$, 
where $\text{Tr}_{n_{0}}$ denotes the trace over the degrees of freedom of constituent $n_{0}$. 
This would however lead to equations of motion where local quantities depend on non-local ones and thus not lead to the same reduction of the complexity of the description as $P_{t}^{n_{0}}$.

Using the projector $P_{t}^{n_{0}}$, see Eq.~(\ref{eq:MoriP-main-text}), we derive an exact equation of motion for the part of the 
density matrix $R(t)$ that is relevant for our interests.
The only assumption made in the derivation is that the initial state factorizes with respect to the considered subsystems, $R(t_{0}) = \bigotimes_{n} \rho_{n} (t_{0})$.
Taking the trace $\text{Tr}_{\not{n_{0}}}$ we find,
\begin{align}\label{reduced NZE main text}
\dot \rho_{n_{0}}(t)=&\mathcal{L}_{n_{0}}\rho_{n_{0}}(t)+\text{Tr}_{\not{n_{0}}} \mathcal{L}_{I} P_{t}^{n_{0}}R(t)\\
&\;+\text{Tr}_{\not{n_{0}}}\mathcal{L}_{I}\int_{t_0}^t dt'\;\mathcal{D}(t,t') \, \mathcal{C}_{t'} \, \mathcal{L}_I P_{t'}^{n_{0}}R(t'), \nonumber
\end{align}
where the action of $\mathcal C_t=\Eins-\sum_{n = 1}^N P_t^n$ is to extract 
correlations and $\mathcal{D}(t,t')=\hat{T} \exp\{\int_{t'}^t dt'' (\mathcal L_0 + \mathcal C_{t''}\mathcal{L}_I)\}$. $\hat T$ orders
any product of operators such that the time arguments increase from right
to left.
The derivation of Eq.~(\ref{reduced NZE main text})
is presented in the Appendix, see \ref{App cMoP} and \ref{App NZE}. 
                                   
Equation~(\ref{reduced NZE main text}) exactly describes the dynamics of reduced density matrices for individual subsystems of a quantum many-body 
system of arbitrary size and geometry. This many-body system may even be open so that its dynamics is not necessarily unitary. Eq.~(\ref{reduced NZE main text}) may 
be viewed as a generalization of the celebrated Nakajima-Zwanzig equation for open systems \cite{Zwanzig01,Breuer07}.
It takes the correlations between subsystems explicitly into account via the action of the projection operator $\mathcal C_t$.
Moreover, whenever the total state $R$ is pure, the growth of the von Neuman entropy of the 
states $\rho_{n}(t)$, as described by Eq.~(\ref{reduced NZE main text}), accounts for the entanglement that is built up between each individual subsystem and its surrounding. 
Since the first line of Eq.~(\ref{reduced NZE main text}) for this case describes a unitary evolution, entanglement between 
subsystems is only taken into account via the second line of Eq.~(\ref{reduced NZE main text}).

Yet, despite only describing the local quantities of the constituent of interest, Eq.~(\ref{reduced NZE main text}) is still very demanding to solve in full generality. 
A viable way for finding its solution is thus to expand it as a Dyson series in powers of the interaction $\mathcal L_I$, see Appendix \ref{App Born}.
Here we keep
terms up to second order in $\mathcal L_I$, which reduces the complexity of the dynamical map $\mathcal D(t,t')$ considerably, and end up with,
\begin{align}\label{NZE in BA main}
  &\dot \rho_{n_{0}}(t)=\mathcal{L}_{n_{0}}\rho_{n_{0}}(t)+\sum_{n=1}^Z \text{Tr}_n \mathcal{L}_{<n_{0},n>}\rho_n(t)\otimes\rho_{n_{0}}(t)\\
&+\sum_{n=1}^Z \text{Tr}_n \mathcal{L}_{<n_{0},n>}\int_{t_0}^t dt'\,\mathcal{K}_{<n_{0},n>}(t,t') \rho_n(t')\otimes\rho_{n_{0}}(t').\nonumber
\end{align}
Here, $\mathcal{K}_{<n_{0},n>}(t,t') = e^{(t-t')(\mathcal{L}_n+\mathcal{L}_{n_{0}})} \mathcal C_{t'}^{<n_{0},n>}\mathcal{L}_{<n_{0},n>}$ 
with $\mathcal C_{t'}^{<n_{0},n>}\equiv \Eins-\rho_n(t')\otimes \text{Tr}_n-\rho_{n_{0}}(t')\otimes \text{Tr}_{n_{0}}$. $\mathcal{L}_{<n_{0},n>}$ 
denotes the interaction between subsystems $n_{0}$ and $n$, and $Z$ the coordination number of the lattice.  


Equation~(\ref{NZE in BA main}) is a nonlinear integro-differential equation for the reduced density matrices $\rho_{n}(t)$ of individual subsystems
and can be integrated numerically using standard techniques \cite{Intdiff}. In particular for large systems where one can assume translation 
invariance, $\rho_{n}(t) = \rho_{n_{0}}(t)$ for all $n$, Eq.~(\ref{NZE in BA main}) reduces to an equation 
for $\rho_{n_{0}}(t)$ only. For these cases our approach thus achieves a similar reduction of computational complexity as 
Gutzwiller type mean-field calculations. Yet, despite this efficiency it is remarkably more accurate than mean-field as we show for a series of examples below. 
Before discussing applications of Eq.~(\ref{NZE in BA main}) we comment on some of its properties. 

We start by noting that whereas one could derive an exact equation
for $\rho_{n_{0}}$ for any choice of the reference state $\rho_{\not{n_{0}}}$, the quality of the approximation in Eq.~(\ref{NZE in BA main}) depends on the ansatz for $\rho_{\not{n_{0}}}$. 
Here, our time dependent and self-consistent projector, defined in Eq.~(\ref{eq:MoriP-main-text}), appears to be the best ansatz to account for a highly dynamical quantum many-body environment. 
Moreover, a Markov approximation as applied in open system theory is here
not applicable since reduced density matrices of the 'system' and its surrounding are treated on an equal footing and correlation 
functions of the surrounding can thus not be expected to decay faster than the dynamics generated by the coupling between subsystems. 
Consequently, our approach does not require any separation of the time scales.
In fact, for models with
time scale separations, which are the basis of approaches using adiabatic elimination or approximate Schrieffer-Wolff transformations \cite{Cirac92}, Eq. (\ref{NZE in BA main}) 
reduces to a standard master equation \cite{Breuer07}.
 
Finally, the first two terms on the right hand side of Eq.~(\ref{reduced NZE main text}) are equivalent to the
mean-field or Gutzwiller approach which has been exploited with remarkable success in equilibrium physics \cite{Sachdev11,Rokhsar91,Fisher89} and was the starting point 
for recent investigations of non-equilibrium systems \cite{Nissen12,Diehl2010,Jin13}.
Mean-field can thus be understood as an approximation to linear order in $\mathcal{L}_{I}$ for the dynamics of single site reduced density matrices. 
Our theory therefore forms a systematic generalization of mean-field approaches. As the non-Markovian properties, the explicit consideration of 
correlations via the projector $\mathcal{C}_{t}$, and entanglement between subsystems are only present in terms of higher than linear order in $\mathcal L_I$,
Eq.~(\ref{NZE in BA main}) yields a different quality of approximation than mean-field. We will show in the sequel that this is indeed the case.
\begin{figure*}[tbp]
\centering
\includegraphics[width=0.8\textwidth]{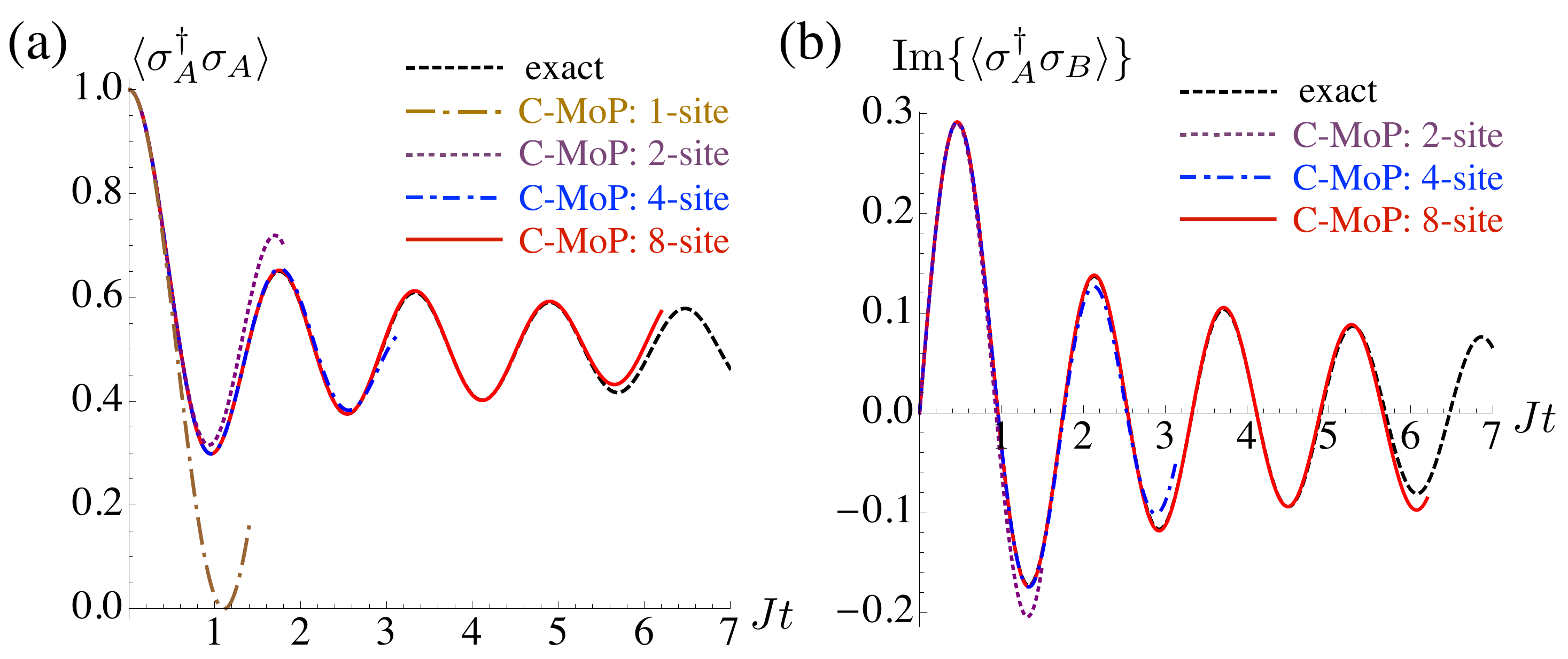}
\caption{ (Color online) Application to unitary dynamics of closed quantum many-body systems. {\bf(a)} $\langle \sigma_{A}^{\dagger} \sigma_{A} \rangle (t)$; exact value (dashed black) \cite{Flesch08}, 
single-site c-MoP (dash-dotted brown), 2-site cluster c-MoP (dotted purple), 4-site cluster c-MoP (dash-dotted blue), and 8-site cluster c-MoP (solid red).
{\bf(b)} $\langle \sigma_{A}^{\dagger} \sigma_{B} \rangle (t)$; exact value (dashed black) \cite{Flesch08}, 2-site cluster c-MoP (dotted purple), 4-site cluster 
c-MoP (dash-dotted blue), and 8-site cluster c-MoP (solid red).
The real part of $\langle \sigma_{A}^{\dagger} \sigma_{B} \rangle (t)$ vanishes for all $t$.}
\label{1Dunitary Fig}
\end{figure*}
\section{Applications and accuracy tests}\label{Sec:tests}
We now turn to test the accuracy of Eq.~(\ref{NZE in BA main}) in applications to one- or two-dimensional systems where either exact solutions or very accurate t-DMRG simulations are available 
for comparison.
In doing so we focus on a lattice of two-level systems or spins and extensions thereof which include coherent drives and relaxation of individual spins. 
This model is described by Eq.~(\ref{EOM main text}) with ($\hbar=1$), 
%
%
\begin{equation}
\mathcal{L}_{n} (\cdot) =-i\,[H_{n},\cdot]+ D_n (\cdot) \: \: \:  \text{and} \: \: \: 
\mathcal{L}_{I} (\cdot) =-i\,[H_{I},\cdot],
\label{full BH dynamics}
\end{equation}
where $H_{n} = \Delta \sigma^{\dagger}_n \sigma_n + (\Omega/2)(\sigma^{\dagger}_n+\sigma_n)$,
$H_{I} = -J\sum_{<n,m>} (\sigma^{\dagger}_n\sigma_m+\sigma_n\sigma^{\dagger}_m)$ 
and $D_n(\cdot)=(\gamma/2) [2 \sigma_n(\cdot)\sigma_n^\dagger-\sigma_n^\dagger \sigma_n(\cdot)-(\cdot)\sigma_n^\dagger \sigma_n]$.
Here, $\sigma_n = \ket{0_{n}}\bra{1_{n}}$ is the de-excitation operator on site $n$.
We have written the Hamiltonian in a rotating frame such that, $\Delta=\omega-\omega_{L}$ is the detuning between spin transition 
frequency $\omega$ and drive frequency $\omega_{L}$. $J$ is the tunneling rate between nearest-neighbor sites indicated by the notation $<n,m>$, and $\Omega$ the drive amplitude.
This model allows us to study both, the unitary dynamics of a closed system as well as stationary states of driven-dissipative systems. 
\subsection{Unitary dynamics of closed systems}\label{Sec:tests closed}
In a first example we consider a one-dimensional closed system version of Eq.~(\ref{full BH dynamics}) with $\omega_{L} = \Omega = \gamma = 0$ and periodic boundary 
conditions that is initially prepared in a pure state with one excitation in every second site and none otherwise,
$\ket{\psi_{0}} = \ket{\dots, 0, 1, 0, 1, \dots}$ \cite{Trotzky12}. 
As this model has an exact solution \cite{Flesch08} we use it to test the accuracy of Eq.~(\ref{NZE in BA main}), see Appendix \ref{App Spin lattice} for its explicit form for the model 
of Eq.~(\ref{full BH dynamics}).

For this homogeneous model with staggered initial conditions all initially occupied sites (denoted A-sites) and all initial empty sites (B-sites) have the 
same reduced density matrices $\rho_{A}(t)$ respectively $\rho_{B}(t)$. We thus denote operators acting on A-sites (B-sites) $\sigma_{A}$ ($\sigma_{B}$).
The exact result for $\langle \sigma_{A}^{\dagger} \sigma_{A} \rangle (t)$ is shown in dashed black in Fig.~\ref{1Dunitary Fig}(a).
For the present setup, Eq.~(\ref{NZE in BA main}) leads to two coupled equations for $\rho_{A}(t)$ and $\rho_{B}(t)$. 
$\text{Tr}\{ \sigma^{\dagger}_{A} \sigma_{A} \rho_{A}(t)\}$, as resulting from this c-MoP calculation is shown in dash-dotted brown in Fig.~\ref{1Dunitary Fig}(a).
One can also group two adjacent spins together and consider the resulting spin dimer as one subsystem described by a reduced density matrix $\rho_{AB}$.
This procedure leads to a cluster version of our approach where all clusters are here initially in the same state $\ket{1_{A},0_{B}}$. 
$\text{Tr}\{ \sigma^{\dagger}_{A} \sigma_{A} \rho_{AB}(t)\}$, as resulting from this cluster c-MoP calculation is shown in dotted purple in Fig.~\ref{1Dunitary Fig}(a).
One can also consider larger clusters, e.g. $\text{Tr}\{ \sigma^{\dagger}_{A} \sigma_{A} \rho_{ABCD}(t)\}$ for 4-site clusters described by $\rho_{ABCD}$ and initially 
prepared in $\ket{1_{A},0_{B}, 1_{C}, 0_{D}}$. 
Results from 4-site and 8-site cluster c-MoP calculations are shown in dashed-dotted blue and solid red in Fig.~\ref{1Dunitary Fig}(a).

We observe that the accuracy of our approach is excellent for short times, but as expected eventually deteriorates for longer times. The same properties can 
be seen for correlations $\langle \sigma_{A}^{\dagger} \sigma_{B} \rangle (t)$ in Fig~\ref{1Dunitary Fig}(b). Remarkably the time range in which the approximation is highly 
accurate grows significantly as one applies the c-MoP approach to increasingly larger clusters. This tendency suggest that even substantially larger time ranges should become 
accessible as one increases the cluster size further. 
To appreciate this perspective one should compare the time ranges that we are able to accurately describe here (although only for local quantities) to those reached with t-DMRG 
approaches ($J t = 6$) on high performance computing clusters \cite{Flesch08}. 
We note that mean-field terms vanish in our example, $\text{Tr}\{\sigma_{n} \rho_{n}(t)\}=0$ for all $t$, due to excitation number conservation.
A mean-field calculation for clusters of $m$ lattice sites will thus be identical to the result for an $m$-site open boundary lattice, and therefore be inaccurate.
\begin{figure*}
\centering
\includegraphics[width=0.8\textwidth]{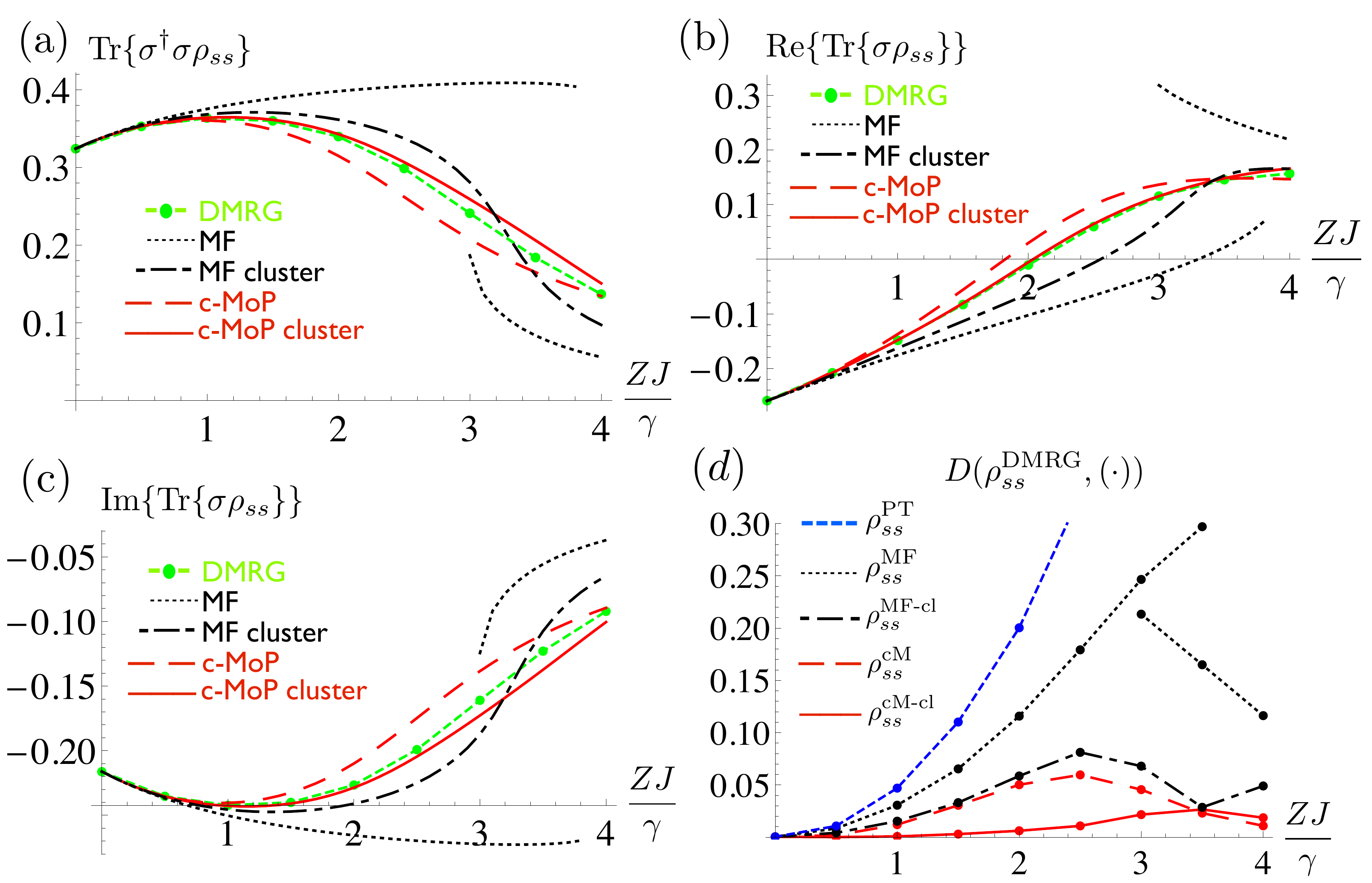}
\caption{\label{DMRG Fig} (Color online) Application to stationary states of driven-dissipative quantum many-body 
systems. {\bf (a)} $\text{Tr}\{\sigma^{\dagger} \sigma \rho_{ss}\}$, {\bf (b)} real part, and {\bf (c)} imaginary part of $\text{Tr}\{\sigma \rho_{ss}\}$ as a 
function of $Z J/\gamma$ for $Z=2$, $\Delta=0.6 \gamma$ and $\Omega=1.5 \gamma$. t-DMRG (dashed green), single-site mean-field (dotted black), two-site cluster mean-field 
(dash-dotted black), single-site c-MoP (dashed red), and two-site cluster c-MoP (solid red).
The bistabilities of single-site mean-field appear to be an artifact of this specific method \cite{Spohn,Schirmer}, see Appendix \ref{App bistability}.
{\bf (d)} Trace distances $D(\rho_{ss}^{\text{DMRG}},\rho_{ss}^{\text{MF}})$ (dotted black), $D(\rho_{ss}^{\text{DMRG}},\rho_{ss}^{\text{MF-cl}})$ 
(dash-dotted black), $D(\rho_{ss}^{\text{DMRG}},\rho_{ss}^{\text{cM}})$ (dashed red), $D(\rho_{ss}^{\text{DMRG}},\rho_{ss}^{\text{cM-cl}})$ (solid red), 
and $D(\rho_{ss}^{\text{DMRG}},\rho_{ss}^{\text{PT}})$ (blue) for the same parameters as (a).}
\end{figure*}
\subsection{Stationary states of driven-dissipative systems}\label{Sec:tests open}
Instead of numerically integrating Eq.~(\ref{NZE in BA main}) we now focus on the physically very interesting scenario of steady states in driven-dissipative 
quantum many-body systems \cite{Barreiro11,Prosen08,Hartmann10,Nissen12,Diehl2008,Diehl2010,Diehl11,Rivas12}. 
We thus consider the model~(\ref{full BH dynamics}) with $\Omega \ne 0$ and $\gamma \ne 0$, assume translation invariance and periodic boundary 
conditions such that $\rho_{n}(t) = \rho_{n_{0}}(t)$ for all $n$, and drop site indices in the following. Equation~(\ref{NZE in BA main}) can be simplified 
significantly if one is only interested in the steady state solution $\rho_{ss} = \lim_{t \to \infty } \rho_{n_{0}}(t)$ since the 
action of the integral kernel $\mathcal{K}_{<n_{0},n>}(t,t')$ on $ \rho_n(t')\otimes\rho_{n_{0}}(t')$ vanishes for $|t-t'|$ sufficiently large. 
For $t\to\infty$ one can thus approximate $\rho_{n}(t') \approx \rho_{ss}$ in the right hand side of Eq.~(\ref{NZE in BA main}) and extract an algebraic equation for $\rho_{ss}$
(see Appendix \ref{appendix steady state}),
\begin{equation}
 0=\left(\mathcal{L}_{\text{LT}} + \mathcal{L}_{\text{MF}}^{ss}+\mathcal{L}_{\text{BT}}^{ss}\right)\rho_{ss},
 \label{eq:c-MoP-main-text}
\end{equation}
where $ \mathcal{L}_{\text{LT}} \propto J^{0}$, $\mathcal{L}_{\text{MF}}^{ss} \propto Z J^{1}$ and $\mathcal{L}_{\text{BT}}^{ss} \propto Z J^{2}$. The explicit expressions for these 
time independent superoperators are given in the Eq.~(\ref{SS zeroth SI},\ref{SS MF SI},\ref{SS Born SI},\ref{corr func ss}). Since $\mathcal{L}_{\text{MF}}^{ss}$ and $\mathcal{L}_{\text{BT}}^{ss}$ depend on $\rho_{ss}$, 
Eq.~(\ref{eq:c-MoP-main-text}) is a nonlinear algebraic equation for $\rho_{ss}$, the roots of which can be found numerically with standard routines.
One can follow the same procedure for subsystems formed by clusters of two, three or more adjacent lattice sites, which results in a cluster c-MoP calculation.  

We test Eq.~(\ref{eq:c-MoP-main-text}) for subsystems consisting of single sites and two-site clusters by comparing its solution, for which we
denote single-site reduced density matrices by $\rho_{ss}^{\text{cM}}$ and $\rho_{ss}^{\text{cM-cl}}$ respectively, to t-DMRG integrations of Eq.~(\ref{EOM main text}) for the one-dimensional 
model in Eq.~(\ref{full BH dynamics}) with $N=21$ lattice sites and open boundary conditions \cite{Hartmann10,Hartmann09}.
From this t-DMRG numerics, which integrated Eq.~(\ref{EOM main text}) for a time range $T = 20/\gamma$ using a second order Trotter expansion with 
steps $\delta t = 10^{-3}/\gamma$, we extract the reduced density matrix for the central site $n_{0}=11$, denoted by $\rho_{ss}^{\text{DMRG}}$.
For comparison we also consider mean-field results for single sites and two-site clusters, denoted by $\rho_{ss}^{\text{MF}}$ and $\rho_{ss}^{\text{MF-cl}}$ respectively, 
and results of standard perturbation theory to second order in the interactions \cite{delValle13,Li13}, denoted by $\rho_{ss}^{\text{PT}}$, to show 
the dramatic quantitative and qualitative improvement of c-MoP over these approaches. For this purpose we compare expectation values of on-site observables and calculate 
the trace distance $D(\rho_1,\rho_2)=\frac{1}{2}|\rho_1-\rho_2|$ with $|A|=\sqrt{A A^\dagger}$ \cite{Nielsen} 
between the t-DMRG result, $\rho_{1} = \rho_{ss}^{\text{DMRG}}$, and the 
approximations, $\rho_{2} = \rho_{ss}^{\text{cM}}, \rho_{ss}^{\text{cM-cl}}, \rho_{ss}^{\text{MF}}$, $\rho_{ss}^{\text{MF-cl}}$ or $\rho_{ss}^{\text{PT}}$.

Figure~\ref{DMRG Fig}(a) shows the occupation number $\text{Tr}\{\sigma^{\dagger} \sigma \rho_{ss}\}$, whereas Figs.~\ref{DMRG Fig}(b) and (c) show the real and imaginary 
parts of $\text{Tr}\{\sigma\rho_{ss}\}$ for t-DMRG, c-MoP and mean-field calculations.
We find a very good agreement between the c-MoP results (red) and t-DMRG results (green), which again improves significantly for two-site clusters (solid red lines) 
compared to individual lattice sites (dashed red lines).
The mean-field results however deviate from the t-DMRG results to an extend which
makes them unreliable over a large parameter range for both, single-site (black dotted) as well as two-site cluster (black dash-dotted) versions. 
These findings are further illustrated by Fig.~\ref{DMRG Fig} (d) which shows the distances $D(\rho_{ss}^{\text{DMRG}},\rho_{ss}^{\text{MF}})$ 
(dotted black), $D(\rho_{ss}^{\text{DMRG}},\rho_{ss}^{\text{MF-cl}})$ (dash-dotted black), $D(\rho_{ss}^{\text{DMRG}},\rho_{ss}^{\text{cM}})$ 
(dashed red), $D(\rho_{ss}^{\text{DMRG}},\rho_{ss}^{\text{cM-cl}})$ (solid red), and $D(\rho_{ss}^{\text{DMRG}},\rho_{ss}^{\text{PT}})$. 
%
%
Here our t-DMRG calculations required bond dimensions up to $400$ which indicates that substantial correlations between subsystems are accurately taken into account by Eq. (\ref{eq:c-MoP-main-text}).

%
\subsection{Comparison of accuracy for stationary states of one- and two-dimensional lattices}\label{Sec:tests 2d}

To further elucidate the versatility of our approach, we here examine its accuracy for stationary states of two-dimensional lattices in comparison to one-dimensional chains. 
\begin{figure*} 
\includegraphics[width=0.8\textwidth]{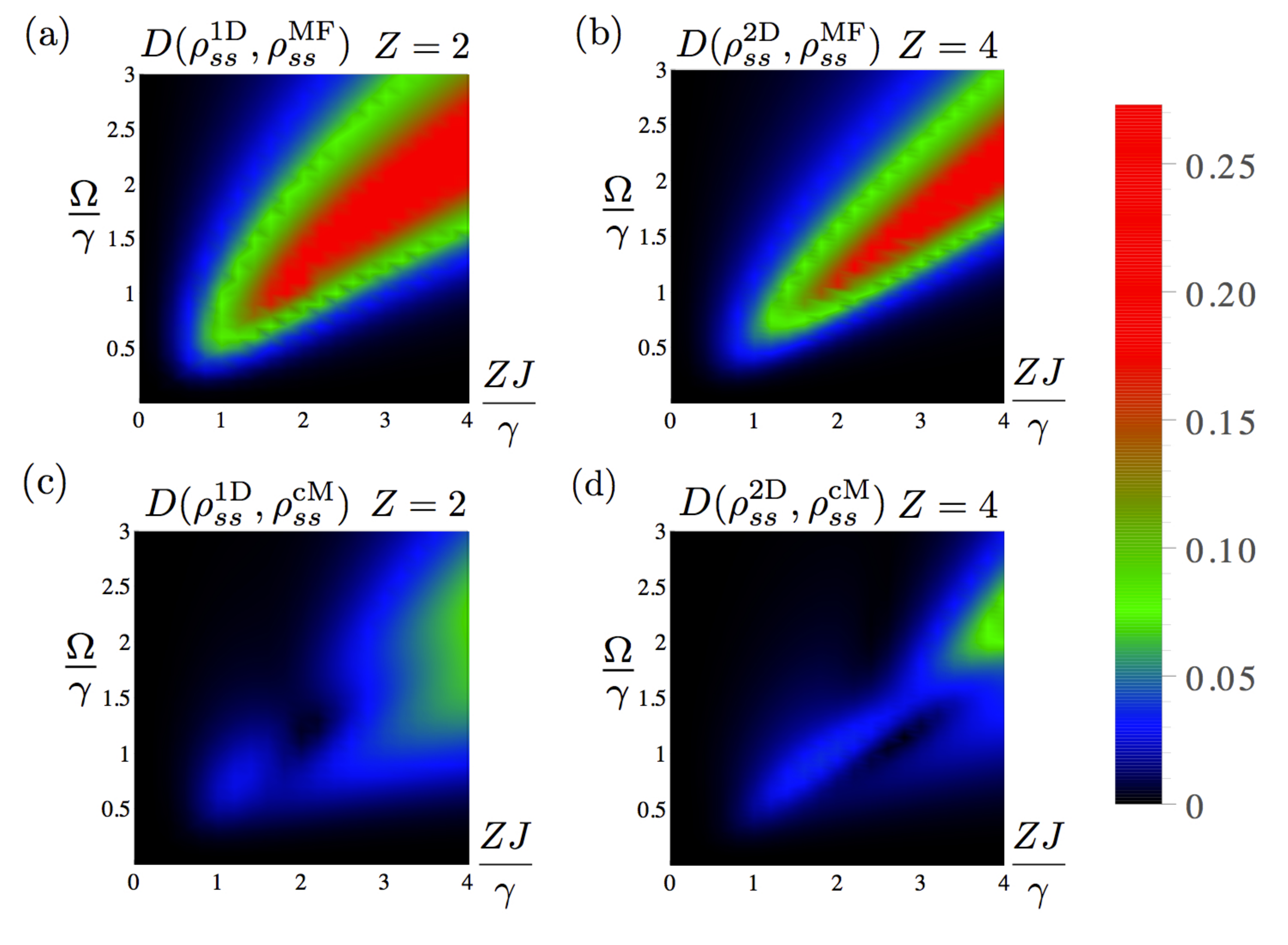}
\caption{\label{1D2D Fig} (Color online) Performance of the method in terms of trace distances from exact solutions for small systems.
(a) and (c): $D(\rho_{ss}^{\text{1D}},\rho_{ss}^{\text{MF}})$ and $D(\rho_{ss}^{\text{1D}},\rho_{ss}^{\text{cM}})$ respectively for $Z=2$, $\Delta=0.5\gamma$, and  $N=3$ with 
periodic boundary conditions as functions of $ZJ/\gamma$ and $\Omega/\gamma$.
(b) and (d): $D(\rho_{ss}^{\text{2D}},\rho_{ss}^{\text{MF}})$ and $D(\rho_{ss}^{\text{2D}},\rho_{ss}^{\text{cM}})$ respectively for $Z=4$, $\Delta=0.5\gamma$, and $N=5$ with 
periodic boundary conditions as functions of $ZJ/\gamma$ and $\Omega/\gamma$.
In the bistable regions of the mean-field approximation we have chosen the branch which is closer to the exact solution.}
\end{figure*}
Fig.~\ref{1D2D Fig} compares the solutions of single-site c-MoP approach, a single-cite mean-field approach and a numerically exact approach for small lattices of 
one ($Z=2$) and two ($Z=4$) dimensions. Trace distances between mean-field approximations and exact solutions are plotted in the upper row whereas the lower row shows 
trace distances between c-MoP approximations and exact solutions.
Figs.~\ref{1D2D Fig}(a) and (c) show $D(\rho_{ss}^{\text{1D}},\rho_{ss}^{\text{MF}})$ and $D(\rho_{ss}^{\text{1D}},\rho_{ss}^{\text{cM}})$ respectively for $Z=2$, $\Delta=0.5\gamma$, 
and  $N=3$ with periodic boundary conditions as functions of $ZJ/\gamma$ and $\Omega/\gamma$.
Figs.~\ref{1D2D Fig}(b) and (d) in turn show $D(\rho_{ss}^{\text{2D}},\rho_{ss}^{\text{MF}})$ and $D(\rho_{ss}^{\text{2D}},\rho_{ss}^{\text{cM}})$ respectively 
for $Z=4$, $\Delta=0.5\gamma$, and $N=5$ with periodic boundary conditions as functions of $ZJ/\gamma$ and $\Omega/\gamma$.
The lattice sizes $N=3$ in one dimension and $N=5$ in two dimensions are chosen because these are the minimal lattice sizes where each 
lattice site has distinct left and right neighbors which close the lattice in periodic boundary conditions in each dimension. Nonetheless 
both lattices are small enough to allow for full numerical solutions for their stationary states.  

Firstly, we notice that there is no bistability for 
$\rho_{ss}^{\text{cM}}$ in the whole parameter range. In the bistable regions of the mean-field approximation we have here chosen the branch 
which is closer to the exact solution. Secondly, we find a remarkable quantitative improvement of c-MoP over mean-field especially for regions 
where the on-site parameter $\Omega$ is comparable to the tunneling, i.e. $\Omega\approx ZJ$. For $\Omega \gg Z J$ both approximations become very good as the dynamics is
dominated by the on-site Liouvillian $\mathcal{L}_{\text{LT}}$. In the opposite case of $Z J \gg \Omega$ the steady state of Eq.~(1) of the main text is close to the 
vacuum state which is here a product state leading to high accuracy for both approximations.
Finally, we find that both approaches become more accurate for a two dimensional lattice, see Fig.~\ref{1D2D Fig}(b) and (d), where mean-field however still remains unsatisfactory. 
\section{conclusions and outlook} \label{conclusions}
In conclusion we have derived an exact equation of motion for the reduced density matrices of subsystems of quantum many-body systems. When expanded in powers of the interaction between 
subsystems or clusters of subsystems, our equation leads to a highly efficient and very accurate approximation of the dynamics of local quantities.
Although only of the same computational complexity, our equation is a significant qualitative and quantitative improvement of mean-field approaches and can also be straightforwardly 
extended to describe clusters of subsystems. 
The accuracy of the approach improves very fast as the size of the clusters is increased and the convergence of the results with increasing cluster size is a control 
handle for verifying their reliability.
The method gives access to correlations of lengths below or equal to the cluster size but an extension to the calculation of longer range correlations is 
straightforward since the considered clusters can also be composed of non-adjacent subsystems.

For unitary dynamics it is capable of covering time ranges comparable to those accessible with cutting edge t-DMRG calculations for small cluster sizes already.
Moreover it takes into account additional terms that appear to be highly relevant in two-dimensional lattices, which makes it a highly promising candidate for the description of 
these systems and their still illusive physics.
For steady states, it reduces to a simple algebraic equation that provides a promising technique for exploring phase diagrams of driven-dissipative systems. 
When combined with techniques developed in \cite{delValle12} it will allow to efficiently compute output spectra of photonic quantum many-body systems.
Some intriguing tasks for future research would be to extend the c-MoP approach to larger size clusters and to investigate higher order terms in the expansion of Eq.~(\ref{reduced NZE main text}).

%
\acknowledgments
The authors thank Elena del Valle, Robert Jirschik, Frank Glowna, Wilhelm Zwerger and Martin Plenio for discussions and comments.
This work was supported by the German Research Foundation (DFG) via the Emmy Noether grant HA 5593/1-1 and the CRC 631.
%
%
\appendix

\section{Time dependent Mori projector and its relation to the theory of open quantum systems}\label{App cMoP}
We consider a physical setup with an N-partite structure, where we are only interested in the degrees of freedom 
of one subsystem of the full setup. Our approach is inspired by the theory of open quantum systems. We thus refer to our subsystem of interest as the 
"system" and to the remaining $N-1$ subsystems as the "environment". The latter
will be traced out in our considerations. The dynamics of the entire setup is described by Eq.~(\ref{EOM main text}) of the main text, which we restate here for completeness,  
\begin{equation}\label{EOM}
  \begin{split}
\dot R(t)&=\mathcal{L}R(t)=\left(\sum_{n=1}^N\mathcal{L}_{n}+\mathcal{L}_{I}\right)R(t) \\
&\equiv \left(\mathcal{L}_{n_{0}}+\mathcal{L}_{\not{n_0}}+\sum_{n=1}^{Z}\mathcal{L}_{<n_{0},n>}+
\mathcal{L}_{I\not{n_0}}\right)R(t).
\end{split}
\end{equation}
To make the relation to open quantum system approaches more transparent, we have here grouped the superoperators into four parts. One part given by $\mathcal{L}_{n_{0}}$ only acts on 
the ``system'', i.e. the subsystem of interest. Another part $\mathcal{L}_{\not{n_{0}}}\equiv \sum_{n=1,n\neq n_0}\mathcal{L}_n$ which describes 
the dynamics of each of the remaining $N-1$ subsystems in
the environment of the subsystem of interest. And finally, two parts contained in the superoperator $\mathcal{L}_I$, where $\sum_{n=1}^{Z}\mathcal{L}_{<n_{0},n>}$ denotes the pairwise interaction 
of the subsystem
$n_0$ with $Z\leq N-1$ different subsystems $n$, and where $\mathcal{L}_{I\not{n_0}}$ accounts for any interaction between subsystems excluding the subsystem of interest. By assigning 
$\mathcal{L}_{I\not{n_0}}$
to the interaction part $\mathcal{L}_I$, we treat all $N$ subsystems on an equal footing.
Moreover, we do not only consider unitary dynamics, where $\mathcal{L}(\cdot)=-i[H,(\cdot)]$. Nevertheless, all superoperators shall be the generators of completely positive, 
trace preserving maps of Lindblad type \cite{Lindblad}.

As we are only interested in ``system'' observables, it is sufficient to know the reduced density matrix of subsystem $n_{0}$ given by $\rho_{n_{0}}(t)=\text{Tr}_{\not{n_{0}}} \{R(t)\}$.
The guiding idea of our approach is thus to introduce a projector $P$, similar to the Mori projector \cite{Mori65,Breuer07}, 
which projects the full density matrix onto a relevant fraction $R_{rel}(t)=PR(t)$ with $P(\cdot)=\rho_{\not{n_{0}}} \otimes \text{Tr}_{\not{n_{0}}}(\cdot)$. 
In strong contrast to open system theory, we however introduce a time dependent projection operator defined in Eq.~(\ref{eq:MoriP-main-text}) of the main text, that is,
\begin{equation}\label{Mori P}
  P_t^{n_{0}}(\cdot)=\rho_{\not{n_{0}}}(t)\otimes \text{Tr}_{\not{n_{0}}}(\cdot),
\end{equation}
where the density matrix $\rho_{\not{n_{0}}}(t)$ is given by a factorized state of the reduced density matrices of all $N-1$ environmental constituents, i.e.
\begin{equation}\label{dens matr env}
\rho_{\not{n_{0}}}(t)=\bigotimes_{n\not{=} n_{0}}\rho_n(t) \;\;\text{with}\;\;\rho_n(t)=\text{Tr}_{\not{n}} R(t).
\end{equation}
The term `relevant` indicates that $PR(t)$ contains all information needed to determine
the exact expectation value of any system operator $A_{n_{0}}$, i.e. $\Bra A_{n_{0}}\Ket(t)=Tr\{A_{n_{0}}PR(t)\}$. 
We emphasize that the environmental density matrix 
$\rho_{\not{n_{0}}}(t)$ or rather each reduced matrix $\rho_n(t)$ is determinded consistently from the evolution $R(t)$ of the full physical setup. Thus, we speak of a self-consistent
Mori projector approach as both the reduced density matrix of the system $\rho_{n_{0}}(t)$ and the environmental state $\rho_{\not{n_{0}}}(t)$ are 
determined consistently with the full dynamics given by $R(t)$.
%
%
Similar to standard open system theory our derivation also employs the complement of $P_t^{n_{0}}$ which projects out the irrelevant part of 
the density matrix $R_{irr}(t)=Q_t^{n_{0}}R(t)$ and is given by
\begin{equation}\label{Q Proj}
Q_t^{n_{0}}(\cdot)=\Eins-P_t^{n_{0}}(\cdot),
\end{equation}
where $\Eins$ is the identity mapping. As usual, we find the complementarity of the two subspaces $R_{rel}(t)$ and $R_{irr}(t)$, i.e.
$P_t^{n_{0}}Q_t^{n_{0}}=Q_t^{n_{0}}P_t^{n_{0}}=0$ and $\Eins=P_t^{n_{0}}+Q_t^{n_{0}}$, and both projectors share the 
characteristic features $(P_t^{n_{0}})^2 = P_t^{n_{0}}$ and $(Q_t^{n_{0}})^2 = Q_t^{n_{0}}$.

\section{Exact Nakajima-Zwanzig type equation for one subsystem}\label{App NZE}
To derive an exact equation of motion for one subsystem, we first derive a set of equations for the two complements $P_t^{n_{0}}R(t)$ and $Q_t^{n_{0}}R(t)$ of the 
full density matrix $R(t)$, then state a formal solution for the irrelevant part and finally deduce a closed equation for the relevant part. For time independent 
projectors, the analog of this equation is known as the Nakajima-Zwanzig equation \cite{Zwanzig01,Breuer07}.
On our way we will exploit the identity $\Eins=P_t^{n_{0}}+Q_t^{n_{0}}$, the full dynamics given by Eq.~(\ref{EOM}) and the relation $\dot P_t^{n_{0}} Q_t^{n_{0}}=0$.
We start with the equation of motion for the relevant fraction of the full density matrix
\begin{align}\label{R_rel EOM SI}
  \dot R_{rel}&=\dot{P}_t^{n_{0}}R(t)+P_t^{n_{0}}\dot{R}(t)= \dot{P}_t^{n_{0}}\Eins R(t)+P_t^{n_{0}} \mathcal{L}\Eins R(t) \nonumber \\
&=\dot P_t^{n_{0}}(P_t^{n_{0}} + Q_t^{n_{0}})R(t)+P_t^{n_{0}}\mathcal{L}(P_t^{n_{0}} + Q_t^{n_{0}})R(t)\nonumber \\
&=\left(\dot P_t^{n_{0}}+P_t^{n_{0}}\mathcal{L}\right)P_t^{n_{0}}R(t)+P_t^{n_{0}}\mathcal{L}Q_t^{n_{0}}R(t).
\end{align}
In contrast to standard open system theory we do not only have a formal time dependence in the projectors but also a new term 
$\dot P_t^{n_{0}} P_t^{n_{0}}R(t)$ arising from the explicit time dependence of the environmental state $\rho_{\not{n_{0}}}(t)$. Next, we use the operator equality 
$\dot Q_t^{n_{0}}=-\dot P_t^{n_{0}}$ to obtain an equation of motion for the irrelevant part
of the full density matrix. In analogy to Eq.~(\ref{R_rel EOM SI}) we find
\begin{equation}\label{R_irr EOM SI}
 \begin{split}
  \dot R_{irr}=\left(-\dot P_t^{n_{0}}+Q_t^{n_{0}}\mathcal{L}\right)P_t^{n_{0}}R(t)+Q_t^{n_{0}}\mathcal{L}Q_t^{n_{0}}R(t).
 \end{split}
\end{equation}
We proceed with the treatment of the environmental density matrix and its time derivative. Therefore, we employ the properties of a trace preserving generator
given for all superoperators from Eq.~(\ref{EOM}). In particular, we will use that $\text{Tr}_n \mathcal{L}_n (\cdot)=0,\;\forall n$ and find
\begin{equation}
\begin{split}\label{rhoEdot SI}
 \dot P_t^{n_{0}} &P_t^{n_{0}}R(t)=\dot \rho_{\not{n_{0}}}(t)\otimes\rho_{n_{0}}(t)=\sum_{m\not{=} n_{0}}\dot \rho_m(t)\otimes\bigotimes_{n \neq m} \rho_n(t)\\
&=\sum_{m\not{=} n_{0}} \text{Tr}_{\not m} \{\dot R(t)\}\otimes\bigotimes_{n \neq m}\rho_n(t)\\
&=\sum_{m\not{=} n_{0}} \text{Tr}_{\not m} \left\{\left(\sum_{j=1}^N\mathcal{L}_j+\mathcal{L}_I\right) R(t)\right\}\otimes\bigotimes_{n \neq m} \rho_n(t)\\
&=\sum_{m\not{=} n_{0}} \left[\text{Tr}_{\not m} \{\mathcal{L}_I R(t)\}+\mathcal{L}_m\rho_m(t)\right]\otimes\bigotimes_{n \neq m} \rho_n(t)\\
&\equiv\mathcal{L}_{\not{n_{0}}} P_t^{n_{0}} R(t)+P_t^{\not{n_{0}}}\mathcal{L}_I(P_t^{n_{0}}+Q_t^{n_{0}})R(t),
\end{split}
\end{equation}
where, in analogy to the projector defined in Eq.~(\ref{Mori P}), we have introduced the projector 
\begin{equation}\label{E Proj}
 P^{\not{n_{0}}}_{t}(\cdot)\equiv \sum_{m\not{=} n_{0}}P_t^m(\cdot) = \sum_{m\not{=} n_{0}}\bigotimes_{n \neq m} \rho_n(t)\otimes \text{Tr}_{\not m}\{\cdot\}.
\end{equation}
We observe that for an $N$-partite physical setup we have an ensemble of $N$ projectors $P_t^m$. By picking one part, $n_{0}$, of the full setup as the "system" of interest,
we have picked $P_t^{m=n_{0}}=P_t^{n_{0}}$ as our projector of interest. Now, we find a connection between all $N$ projectors due to the explicit time dependence of the environmental reference
state $\rho_{\not{n_{0}}}(t)$. Note, that the projector $P_t^{\not{n_{0}}}$, see Eq.~(\ref{E Proj}), depends
on $P_t^{n_{0}} R(t)$ via $\rho_{n_{0}}(t)=\text{Tr}_{\not{n_{0}}}\{P_t^{n_{0}} R(t)\}$. However, there is no dependence on $Q_t^{n_{0}}R(t)$. 
This feature allows us to find a closed equation for the relevant part of the density matrix.

To this end we employ $P_t^{n_{0}} \mathcal{L}_{n_{0}}(\cdot)=\mathcal{L}_{n_{0}} P_t^{n_{0}} (\cdot)$
and restate the equation of motion for the irrelevant part of the full density matrix. Starting from Eq.~(\ref{R_irr EOM SI}) we end up with
\begin{equation}\label{R_irr EOM new SI}
 \begin{split}
\frac{d}{dt} (Q_t^{n_{0}}R(t))=\mathcal C_t\mathcal{L}_IP_t^{n_{0}}&R(t)\\
+&\left(\mathcal C_t\mathcal{L}_I + \mathcal L_0\right)Q_t^{n_{0}}R(t),
 \end{split}
\end{equation}
where we utilize the abbreviations $\mathcal L_0=\sum_{n=1}^N \mathcal L_n$ and $\mathcal C_t=-P_{t}^{\not{n_{0}}}+ Q_{t}^{n_{0}}=\Eins -\sum_{n=1}^N P_t^n$.
Here the projector $\mathcal{C}_{t}$ projects onto the correlations contained in the object it is applied on only.
Before we state the formal solution of Eq.~(\ref{R_irr EOM new SI}), it is convenient to introduce a shorthand notation for the time-propagator 
$\mathcal{D}(t,t')=\hat{T} \exp \{\int_{t'}^t dt'' (\mathcal C_{t''}\mathcal{L}_I + \mathcal{L}_0)\}$
including the time-ordering operator $\hat T$ which orders
any product of superoperators such that the time arguments increase from right
to left \cite{Breuer07}.
By iteratively integrating Eq.~(\ref{R_irr EOM new SI}) we are able to cast the formal solution of $Q_t^{n_{0}}R(t)$, for a given state $R(t_0)$ at an initial time $t_0$, into the form
\begin{equation}
\begin{split}
Q_t^{n_{0}} R(t)=\int_{t_0}^t dt'\;\mathcal{D}(t,t')&\mathcal C_{t'}\mathcal{L}_IP_{t'}^{n_{0}}R(t') \\
&+ \mathcal{D}(t,t_0)\underbrace{Q_{t_0}^{n_{0}}R(t_0)}_{=0}.
\end{split}
\end{equation}
In the context of this work, we focus on a physical situation with a factorized initial state, i.e. $Q_{t_0}^{n_{0}}R(t_0)=0$. This
assumption is natural for the driven-dissipative scenarios we here consider as one of the applications of our approach. Whenever the driving is switched on at time $t_0$, the initial state of 
a dissipative quantum many body system is mostly given by the vacuum which is typically a factorized many body state.
We insert the formal solution of Eq.~(\ref{R_irr EOM new SI}) into Eq.~(\ref{R_rel EOM SI}) and trace over the environmental degrees of freedom to obtain the exact equation 
for the reduced density matrix of the system,
\begin{equation}\label{reduced NZE}
 \begin{split}
  \dot \rho_{n_{0}}&(t) =\mathcal{L}_{n_{0}}\rho_{n_{0}}(t)+\sum_{n=1}^Z\text{Tr}_{\not{n_{0}}}\{\mathcal{L}_{<n_{0},n>} P_t^{n_{0}}R(t)\} \,\\
&+\sum_{n=1}^Z\text{Tr}_{\not{n_{0}}}\{\mathcal{L}_{<n_{0},n>}\int_{t_0}^t dt'\;\mathcal{D}(t,t')\mathcal C_{t'}\mathcal{L}_IP_{t'}^{n_{0}}R(t')\},
 \end{split}
\end{equation}
which is a closed equation for the relevant part of the full density matrix and identical to Eq.~(\ref{reduced NZE main text}) in the main text. By tracing over 
the environment the term $\dot P_t^{n_{0}} P_t^{n_{0}}R(t)$ in Eq.~(\ref{R_rel EOM SI}) 
drops out as $\text{Tr}_{\not{n_{0}}}\{\dot P_t^{n_{0}}(\cdot)\}=0$. Note, that we have also replaced $\text{Tr}_{\not{n_{0}}}\mathcal{L}_I(\cdot)$ by 
$\sum_{n=1}^Z\text{Tr}_{\not{n_{0}}}\mathcal{L}_{<n_{0},n>}$. This can be easily understood as $\mathcal{L}_{I\not{n_0}}$, see Eq.~(\ref{EOM}), is the generator of a trace preserving map acting 
solely on the environmental Hilbert space, and hence $\text{Tr}_{\not{n_{0}}} \mathcal{L}_{I\not{n_0}}=0$.
In the case of vanishing system-environment interaction, i.e. $\mathcal{L}_{<n_{0},n>}=0$, we recover the 
free dynamics of the system being completely independent of the environment.
Eq.~(\ref{reduced NZE}) may be viewed as a generalization of the prominent Nakajima-Zwanzig equation for open systems.
In contrast to the standard Nakajima-Zwanzig equation there is however a dependence on $P_t^{n_{0}} R(t)$ contained 
in the integral kernel and the dynamical map $\mathcal{D}(t,t')$. We emphasize that this is a direct consequence of a time dependent or rather a self-consistent Mori projector ansatz.

\section{Expansion and Born Approximation}\label{App Born}

Regardless of the chosen projector the resulting Nakajima-Zwanzig type equation is often exceedingly difficult to solve in 
full generality and approximations are usually needed. Obviously, equation (\ref{reduced NZE}) can be expanded as a Dyson 
series in powers of the system-environment interaction $\mathcal{L}_I$,
%
%
%
 \begin{widetext}
\begin{equation}\label{reduced NZE expand}
\begin{split}
\dot \rho_{n_{0}}(t) &=\mathcal{L}_{n_{0}}\rho_{n_{0}}(t)+\sum_{n=1}^Z\text{Tr}_{\not{n_{0}}} \mathcal{L}_{I} P_t^{n_{0}}R(t)
+ \text{Tr}_{\not {n_0}} \mathcal{L}_I \int_{t_{0}}^t d t'  e^{\mathcal{L}_0(t-t')}\;\mathcal C_{t'}\mathcal{L}_I P^{n_0}_{t'} R(t')+ \sum_{m=3}^{\infty} \mathcal{Y}_{m}\,,
\end{split}
\end{equation}
where the $m$-th order terms read,
 \begin{equation}\label{mth order corr}
 \begin{split}
\mathcal{Y}_{m} = & \text{Tr}_{\not {n_0}} \mathcal{L}_I \int_{t_{0}}^t d t' \int_{t'}^t dt_{m-1} \; e^{\mathcal{L}_0(t-t_{m-1})}\;\mathcal C_{t_{m-1}}\mathcal{L}_I \int_{t'}^{t_{m-1}} dt_{m-2} \; e^{\mathcal{L}_0(t_{m-1}-t_{m-2})}\;\mathcal C_{t_{m-2}}\mathcal{L}_I \\
& \times \dots \times \int_{t'}^{t_{3}} dt_{2} \; e^{\mathcal{L}_0(t_{3}-t_{2})}\;\mathcal C_{t_2} \mathcal{L}_I e^{\mathcal{L}_0(t_{2}-t')}\;\mathcal C_{t'} \mathcal{L}_I P^{n_0}_{t'} R(t') \, ,
 \end{split}
\end{equation}
\end{widetext}
with a time ordering as $t_0\leq t'\leq t_2\leq ...\leq t_{m-1}\leq t$.
In order to understand the physical processes described by the $m$-th order correction, for $m\geq2$, it is convenient to 
read Eq.~(\ref{mth order corr}) from right to left. There are always $m$ different chronologically ordered points in 
time $\{t_0,t_1,...,t_m\}$, with $t_1=t'$ and $t_m=t$. At each point in time $t_j$, for $1\leq j \leq m-1$, we find an interaction 
vertex described by the superoperator $\mathcal C_j \mathcal L_I$. The interaction vertices are linked via interaction-free time evolution $e^{\mathcal L_0 (t_{j+1}-t_j)}$.  At the very right end 
of Eq.~(\ref{mth order corr}) we find the factorized state $P_{t'}^{n_0}R(t')$ or rather $P_{t_1}^{n_0}R(t_1)$ which, by maintaining the 
order in $\mathcal L_I$, can be rewritten 
into $P_{t_1}^{n_0}R(t_1)=e^{\mathcal L_0 (t_{1}-t_0)}P_{t_0}^{n_0}R(t_0)=e^{\mathcal L_0 (t_{1}-t_0)}R(t_0)$. Therefore, the term $\mathcal Y_m$ 
describes a process where the initially factorized state $R(t_0)$
evolves to the first vertex point at $t_1=t'$ where the action of the superoperator $\mathcal L_I$ builds up correlations which are then projected 
out by the action of $\mathcal C_{t_1}$. The resulting operator given by 
$A\equiv \mathcal C_{t_1} \mathcal{L}_I  e^{\mathcal L_0 (t_{1}-t_0)}R(t_0)$ solely contains the correlated part between any constituents of the 
entire $N$-partite setup which have been build up by the action of 
$\mathcal L_I$. For instance, if $\mathcal L_I$ describes nearest-neighbor interactions on a lattice, then the resulting operator $A$ would contain 
all correlations between any nearest-neighbor pair on the entire lattice.

Similarly, this process continues from one point in time $t_j$ to the following one $t_{j+1}$ successively building up correlations until the last 
point in time $t=t_m$ is reached. The interaction vertex at the present 
time $t=t_m$, however, is not described by the superoperator $\mathcal C_t \mathcal L_I$ but rather 
by $\text{Tr}_{\not {n_0}} \mathcal{L}_I\equiv \text{Tr}_{\not {n_0}} P_t^{n_0}\mathcal{L}_I$. Clearly, the superoperator $P_t^{n_0}$ projects out the 
relevant part of the dynamics of the reduced density matrix $\rho_{n_0}(t)$. In summary, we conclude that the $m$-th order corrections for $m\geq2$ 
contain the influence of correlations which arise due to non-Markovian memory effects in the system-environment interaction, where the role of the system 
is taken by the subsystem with constituent number $n_0$. In turn, the terms up to first order in 
$\mathcal L_I$ do neither contain any correlations nor non-Markovian memory effects.  Hence, we expect a large improvement in the quality of the 
approximations by going from first order in $\mathcal L_I$ to second order in $\mathcal L_I$.

Motivated by these insights we apply the so-called Born approximation \cite{Breuer07} which takes all terms up to second order into account.
We are able to directly formulate Eq.~(\ref{reduced NZE}) in Born approximation by dropping all terms proportional to $\mathcal{L}_I$ from the 
exponent of the dynamical map $\mathcal{D}(t,t')$ and finally arrive at an equation to which we will refer as the c-MoP equation, 
\begin{widetext}
\begin{equation}\label{NZE in BA}
 \begin{split}
  \dot \rho_{n_{0}}(t)=\mathcal{L}_{n_{0}}\rho_{n_{0}}(t)&+\sum_{n=1}^Z \text{Tr}_n\{\mathcal{L}_{<n_{0},n>}\rho_n(t)\otimes\rho_{n_{0}}(t)\}\\
&+\sum_{n=1}^Z \text{Tr}_n\mathcal{L}_{<n_{0},n>}\int_{t_0}^t dt'\;\mathcal{D}_{<n_{0},n>}(t,t') \mathcal C_{t'}^{<n_{0},n>}\mathcal{L}_{<n_{0},n>}\rho_n(t')\otimes\rho_{n_{0}}(t'),
 \end{split}
\end{equation}
\end{widetext}
with $\mathcal{D}_{<n_{0},n>}(t,t')\equiv e^{(t-t')(\mathcal{L}_n+\mathcal{L}_{n_{0}})}$ for the dynamical map describing the free evolution of the ``system'' and the $n$-th
constituent, and
$\mathcal C_{t'}^{<n_{0},n>}\equiv \Eins-\rho_n(t')\otimes \text{Tr}_n-\rho_{n_{0}}(t')\otimes \text{Tr}_{n_{0}}$. This equation is identical to 
equation~(\ref{NZE in BA main}) in the main text. Interestingly, in the Born approximation we only find "non-mixing" terms proportional to 
$\mathcal{L}_{<n_{0},n>}\mathcal{L}_{<n_{0},m>} \delta_{n,m}$ and all possible terms containig $\mathcal{L}_{I\not{n_0}}$ vanish as well.
In fact, this is the consequence of a factorized environmental reference state 
, see Eq.~(\ref{dens matr env}). 

\section{Application to a spin lattice}\label{App Spin lattice}

\subsection{Mean-field terms}\label{App MF section}

The right hand side of Eq.~(\ref{NZE in BA}) shows three terms, which are zeroth, first and second order in $\mathcal{L}_{<n_{0},n>}$, respectively. The
zeroth order term $\mathcal{L}_{n_{0}}\rho_{n_{0}}(t)$ denotes the free evolution of the system which becomes exact for cases without interactions between 
the constituents, $\mathcal{L}_{I} = 0$. The first and zeroth order terms 
taken together are equivalent to the well-known
mean-field approximation. For the model given in Eq. (\ref{full BH dynamics}) of the main text 
with $\mathcal{L}_{<n_{0},n>}(\cdot)=-i [H_I,(\cdot)]$ and $H_I= -J (\sigma_{n_{0}} \sigma_n^\dagger+\text{H.c.})$, the 
equation of motion up to first order in $J$ reads
\begin{equation}\label{PL_IP}
\begin{split}
\dot \rho_{n_{0}}(t)&=\mathcal{L}_{n_{0}}\rho_{n_{0}}(t) \\
&\;\;\;\;\;\;\;\,+iJ\,\sum_{n=1}^Z\,[ \sigma_{n_{0}} \text{Tr}_{n}\{\sigma^{\dagger}_n \rho_n(t)\}+\text{H.c.},\rho_{n_{0}}(t)]\\
&=\mathcal{L}_{n_{0}}\rho_{n_{0}}(t) +iJ\,\sum_{n=1}^Z\,[ \sigma_{n_{0}} \Bra \sigma^{\dagger}_n\Ket(t)+\text{H.c.},\rho_{n_{0}}(t)].
 \end{split}
\end{equation}
The same equation can be obtained by a formal replacement $\sigma_{n_{0}} \sigma_n^\dagger\rightarrow \sigma_{n_{0}} \Bra \sigma_n^\dagger\Ket$ in the 
full dynamics given by Eq.~(\ref{EOM}), where the operator-valued interaction $\sigma_{n_{0}} \sigma_n^\dagger$ is 
replaced by a coupling of the system to a classical field $\Phi$ with $\Phi=\Bra \sigma_n^\dagger\Ket$. 
We thus find that for a general setup described by Eq.~(\ref{EOM}) the mean-field approximation can be understood as 
the first two leading terms of an exact Nakajima-Zwanzig type equation with a time dependent and self consistent 
Mori projector and that the second order term goes beyond mean-field implying that the choice of the time dependent Mori projector is well motivated. 

\subsection{Born Terms}
\label{App Born section unitary}

Next, we want to determine the structure of the terms in Eq.~(\ref{NZE in BA}) which are of second order in $\mathcal{L}_{<n_{0},n>}$. In the following, we will refer to these terms
as the Born terms. We start with the two terms originating from the 
integrand $\mathcal{D}(t,t')Q^{n_0}_{t'}\mathcal{L}_IP^{n_0}_{t'}R(t')\rightarrow(\Eins-\rho_n(t')\otimes \text{Tr}_n)\mathcal{L}_{<n_{0},n>}\rho_n(t')\otimes\rho_{n_{0}}(t')$,
see Eq.~(\ref{reduced NZE}) and Eq.~(\ref{NZE in BA}).
By choosing $t_0=0$ and
substituting $\tau=t-t'$, we find for the model given in Eq.~(\ref{full BH dynamics}) of the main text,
\begin{widetext}
\begin{equation}\label{Deg + Sch terms}
 \begin{split}
 \sum_{n=1}^Z \text{Tr}_{\not{n_{0}}}\mathcal{L}_{<n_{0},n>}\int_{t_0}^t dt'\;&\mathcal{D}(t,t')Q^{n_0}_{t'}\mathcal{L}_IP^{n_0}_{t'}R(t') = 
 -J^2\,\sum_{n=1}^Z\,\sum_{j\in\{-,+\}}\int_{0}^t d\tau \,d_{j}^n(\tau,t)\left[\sigma^j_{n_{0}},e^{\tau\mathcal{L}_{n_{0}}}[\sigma_{n_{0}},\rho_{n_{0}}(t-\tau)]\right]+\text{H.c.}\\
&\;\;\;\;-J^2\,\sum_{n=1}^Z\,\sum_{j\in\{-,+\}}\int_{0}^t d\tau \,s^n_{j}(\tau,t)\left[\sigma^j_{n_{0}},e^{\tau\mathcal{L}_{n_{0}}}\rho_{n_{0}}(t-\tau)\sigma_{n_{0}}\right]+\text{H.c.},
 \end{split}
\end{equation} 
with the correlation functions of the environment
\begin{equation}\label{d + s corr func}
 \begin{split}
d_{j}^n(\tau,t)&= \text{Tr}_n\{(\sigma_n^j)^\dagger e^{\tau\mathcal{L}_n} \sigma_n^\dagger\rho_n(t-\tau)\}
- \text{Tr}_n\{(\sigma_n^j)^\dagger e^{\tau\mathcal{L}_n} \rho_n(t-\tau)\} \, \text{Tr}_n\{\sigma_n^\dagger\rho_n(t-\tau)\}\,,\\
s_{j}^n(\tau,t)&=\;\text{Tr}_n\{(\sigma_n^j)^\dagger e^{\tau\mathcal{L}_n} [\sigma_n^\dagger,\rho_n(t-\tau)]\}
 \end{split}
\end{equation}
\end{widetext}
and a sum $\sum_{j\in\{-,+\}}$ which runs over all possible combinations of operators $\sigma^-\equiv \sigma$ and $\sigma^{+}\equiv \sigma^\dagger$.
Note that we have here restricted $\mathcal{L}_{<n_{0},n>}$ to a unitary case which is in general not necessary. 

Apart from the time dependence in the state of the environment $\rho_n(t-\tau)$, these terms appear in the standard open system approach as well.
Within the Born-Markov approximation they generate the standard Lindblad type terms which describe effects like spontaneous emission 
or incoherent thermal pumping. We emphasize that these terms are only effected by correlation functions of purely environmental type. In contrast, the Born term
generated by the actual time dependence of the environmental state, hence by the integrand $P^{\not{n_{0}}}_{t'}\mathcal{L}_I P_t^{n_{0}} R(t)$ in Eq.~(\ref{reduced NZE}), 
displays a dependence on correlation functions of system type
\begin{widetext}
\begin{equation}\label{Hartm terms} 
\begin{split}
-\sum_{n=1}^Z \text{Tr}_{\not{n_{0}}}\mathcal{L}_{<n_{0},n>}\int_{t_0}^t dt'\;\mathcal{D}(t,t')P^{\not{n_{0}}}_{t'}\mathcal{L}_IP^{n_0}_{t'}R(t')
= -iJ^2\,\sum_{n=1}^Z\,\sum_{j\in\{-,+\}}\int_{0}^t d\tau\,h_{j}^n(\tau,t)[\sigma_{n_{0}},e^{\tau\mathcal{L}_{n_{0}}} \rho_{n_{0}}(t-\tau)]+\text{H.c.},
 \end{split}
\end{equation}
\end{widetext}
where we find a correlation function which includes both system and environmental state dependence given by 
\begin{equation}\label{h corr func}
\begin{split}
h_{j}^n(\tau,t) = i\, &\text{Tr}_{n_{0}}\{(\sigma^j_{n_{0}})^\dagger \rho_{n_{0}}(t-\tau)\} \,\times\\
 &\;\;\;\;\;\;\;\;\;\times\text{Tr}_n\{\sigma_n^\dagger e^{\tau\mathcal{L}_n} [\sigma_n^j,\rho_n(t-\tau)]\}.
\end{split}
\end{equation}
We interpret this term as a back-action of the system. It only appears in the case of a time dependent Mori projector and in second order of the coupling, i.e. in $J^2$, indicating that
the physical interaction process has to take a route starting from the system leading over to the environment and back to the system again. 

\section{Steady State equations}
\label{appendix steady state}
The Nakajima-Zwanzig equation has the structure of an integro-differential equation. The change of the density matrix $\dot\rho_{n_{0}}(t)$ depends not only on the current state $\rho_{n_{0}}(t)$
but also on its past history $\rho_{n_{0}}(t-\tau)$. 
Within the Markov approximation the integral kernel or rather the environmental 
correlation function can be very well approximated by a temporal delta function $\delta(t-\tau)$ resolving the integro-differential structure.
In our approach however, the reduced density matrices of the system and the environment are treated on an equal footing and correlation functions of the environment can not be expected to decay 
faster than the dynamics generated by the system-environment coupling which renders a Markov approximation inappropriate. 

Under certain conditions, however, the equation for the steady state of the density matrix, i.e. 
$\rho_{n_{0}}^{ss}\equiv \lim_{t \to \infty}\rho_{n_{0}}(t)$, does not exhibit the integro-differential structure. We emphasize that, so far, these are the first
restrictions that we apply in the time-dependent and self-consistent Mori projector ansatz. We assume that the superoperator $\mathcal{L}_{\not{n_{0}}}$ or rather $\mathcal{L}_n$ shall describe 
a master equation in Lindblad form instead of a unitary evolution. 
Furthermore, it shall have a unique steady state to which all states relax in the limit $t\to\infty$. For the model given in Eq.~(\ref{full BH dynamics}) of the main text, 
the existence of such a unique steady 
state for a truncated Hilbert space follows directly from Spohn's theorem \cite{Spohn, Schirmer}.  
Then, if such a unique steady state exists, the dynamical map $e^{\tau\mathcal{L}_n}$ asymptotically maps all operators 
onto the same operator for each trace class, i.e. $\lim_{\tau\to\infty} e^{\tau\mathcal{L}_n} A=\lim_{\tau\to\infty} e^{\tau\mathcal{L}_n} B$ 
for all operators $A$ and $B$ with $\text{Tr} \{A\}=\text{Tr} \{B\}$. This implies that all commutators $[A,B]$ vanish under the action of the dynamical map $e^{\tau\mathcal{L}_n}$ in the 
long time limit since $\lim_{\tau\to\infty} e^{\tau\mathcal{L}_n} AB=\lim_{\tau\to\infty} e^{\tau\mathcal{L}_n} BA$. Therefore, we can conclude 
that the correlation functions $s(\tau,t-\tau)$ and $h(\tau,t-\tau)$ given in Eq.~(\ref{d + s corr func}) and 
Eq.~(\ref{h corr func}), respectively, vanish to zero in the limit $\tau\to\infty$ for all $t\geq0$. The correlation function $d(\tau,t-\tau)$ in Eq.~(\ref{d + s corr func})
vanishes as well in this limit, as it can be understood as the correlated part of a two-time correlation function. That is, it is of 
the form $\Bra A_{\not{n_{0}}}(t-\tau)A_{\not{n_{0}}}(t)\Ket - \Bra A_{\not{n_{0}}}(t-\tau) \Ket \Bra A_{\not{n_{0}}}(t)\Ket$ which vanishes for $\tau \to \infty$ for relaxing 
systems that have a unique steady state \cite{Rivas12}.

The physical picture behind the assumption of a relaxing system is that the ``environment'' itself is an open system coupled to a Markovian bath. Intuitively, it can be understood that the 
memory of such an
environment has a finite range into the past because all the information that reaches the Markovian bath is lost forever. Practically, it means that the integral kernels in all the Born terms
vanish to zero for large values of $\tau$. Hence, let $t^{*}$ be the time after which the $\rho_{n}(t)$ reach their steady state such 
that $\rho_{n}(t-\tau) \approx \rho_{n}^{ss}$ for $t-\tau > t^{*}$, and let $\tau^{*}$ be the time for which $s(\tau,t-\tau)$, $h(\tau,t-\tau)$  and $d(\tau,t-\tau)$ have decayed 
to zero. Provided we chose $t > t^{*} + \tau^{*}$ and assume the existence of a steady state, which does not necessarily need to be unique, we can replace $\rho_{n}(t-\tau)\to\rho_{n}^{ss}$ 
for all $n$ in Eq.~(\ref{NZE in BA}). In less mathematical terms, this is to say that deep in the steady state the finite range of the environmental memory can only see the steady state itself.
Taking the limit $t\to\infty$ we thus extract an algebraic equation for the steady state $\rho_{n}^{ss}$ from Eq.~(\ref{NZE in BA}) by replacing 
$\rho_{n}(t-\tau)\to\rho_{n}^{ss}$ for all $n$ in the limit $t\to\infty$ for all $\tau$ with $0\leq\tau<t$. We obtain,
\begin{equation}\label{SS NZE SI}
 0=\,\mathcal{L}_{n_{0}}\rho_{n_{0}}^{ss}+\mathcal{L}_{\text{MF}}^{ss}\rho_{n_{0}}^{ss}+\mathcal{L}_{\text{BT}}^{ss}\rho_{n_{0}}^{ss},
\end{equation}
with $\mathcal{L}_{n_{0}}\rho_{n_{0}}^{ss}$ describing the free evolution of the system, the mean-field term $\mathcal{L}_{\text{MF}}^{ss}\rho_{n_{0}}^{ss}$ 
which is the first order correction in the system-environement coupling strength $J$,
\begin{equation}\label{SS MF1 SI}
 \mathcal{L}_{\text{MF}}^{ss}\rho_{n_{0}}^{ss}=\,iJ\,\sum_{n=1}^Z\,[ \sigma_{n_{0}} \,\text{Tr}_n\{\sigma^{\dagger}_n\rho_n^{ss}\}+\text{H.c.},\rho_{n_{0}}^{ss}],
\end{equation}
and the Born terms beyond mean-field with steady-state dependent correlation functions,
  \begin{widetext}
\begin{equation}\label{SS Born1 SI}
 \begin{split}
\mathcal{L}_{\text{BT}}^{ss}\rho_{n_{0}}^{ss}=&-J^2\,\sum_{n=1}^Z\,\sum_{{j\in\{-,+\}}}\int_{0}^\infty d\tau \,d_{j}^n(\tau,\rho_n^{ss})\left[\sigma^j_{n_{0}},e^{\tau\mathcal{L}_{n_{0}}}[\sigma_{n_{0}},\rho_{n_{0}}^{ss}]\right]+\text{H.c.}\\
&-J^2\,\sum_{n=1}^Z\,\,\sum_{{j\in\{-,+\}}}\int_{0}^\infty d\tau \,s^n_{j}(\tau,\rho_{n}^{ss})\left[\sigma^j_{n_{0}},e^{\tau\mathcal{L}_{n_{0}}}\rho_{n_{0}}^{ss}\sigma_{n_{0}}\right] +\text{H.c.}\\
&-iJ^2\,\sum_{n=1}^Z\,\,\sum_{{j\in\{-,+\}}}\int_{0}^\infty d\tau\,h_{j}^n(\tau,\rho_{n_{0}}^{ss},\rho_n^{ss})\left[\sigma_{n_{0}},e^{\tau\mathcal{L}_{n_{0}}} \rho_{n_{0}}^{ss}\right]+\text{H.c.}.
 \end{split}
\end{equation}
\end{widetext}
Eq.~(\ref{SS NZE SI}) is not closed, yet, as it still depends on the environmental steady state $\rho_n^{ss}$, which is in strong contrast to standard open system theory \cite{Breuer07}, 
not an a priori given state. At this point, however, one could switch the part
of the "system" and the "environment" and obtain two coupled but closed algebraic equations. Or alternatively, one could apply the self-consistency condition 
$\rho_n\cong\rho_{n_{0}}$ whenever it can be justified for the physical situation under study.

Setting $\rho_{n}^{ss} \equiv \rho_{ss}$ and dropping site indices, $\sigma_{n} \equiv \sigma$ and $\mathcal{L}_{n} \equiv \mathcal{L}_{\text{LT}}$, for all $n$ we thus find the 
nonlinear algebraic equation (7) of the main text, where the individual terms read, 
\begin{align}\label{SS zeroth SI}
\begin{split}
\mathcal{L}_{\text{LT}} \rho_{ss} = -i\,&\left[\Delta \sigma^{\dagger} \sigma+\frac{\Omega}{2}(\sigma^\dagger+\sigma),\rho_{ss}\right]\\
&\;\;\;+\frac{\gamma}{2} \left( 2 \sigma\rho_{ss}\sigma^\dagger-\sigma^{\dagger} \sigma\rho_{ss}-\rho_{ss}\sigma^{\dagger} \sigma\right)
\end{split}
\end{align}
for the local terms of order $J^{0}$,
\begin{equation}\label{SS MF SI}
 \mathcal{L}_{\text{MF}}^{ss}\rho_{ss}=\,iZJ\,[ \sigma \,\text{Tr}\{\sigma^\dagger\rho_{ss}\}+\text{H.c.}\,,\rho_{ss}],
\end{equation}
for the mean-field terms of order $J^{1}$, and,
%
\begin{equation}\label{SS Born SI}
 \begin{split}
&\mathcal{L}_{\text{BT}}^{ss}\rho_{ss}=\left\{-iZ J^2\sum_{j}\int_{0}^\infty d\tau h_{j}(\tau,\rho_{ss})\left[\sigma,e^{\tau\mathcal{L}_{\text{LT}}} \rho_{ss}\right]\right.\\
&\left.-Z J^2\sum_{j}\int_{0}^\infty d\tau \,s_{j}(\tau,\rho_{ss})\left[\sigma^{j},e^{\tau\mathcal{L}_{\text{LT}}}\rho_{ss}\,\sigma\right]\right.\\
&\left.-ZJ^2\sum_{j}\int_{0}^\infty d\tau \,d_{j}(\tau,\rho_{ss})\left[\sigma^{j},e^{\tau\mathcal{L}_{\text{LT}}}[\sigma,\rho_{ss}]\right]\right\}   +\text{H.c.}
 \end{split}
\end{equation}
%
for the Born terms of order $J^{2}$.
The sum $\sum_{j}$ with $j\in\{-,+\}$ runs over all possible combinations of operators $\sigma^-\equiv \sigma$ and $\sigma^{+}\equiv \sigma^\dagger$. Moreover, the steady-state 
dependent correlation functions
are given by
\begin{equation}\label{corr func ss}
 \begin{split}
h_{j}(\tau,\rho_{ss})=&\,i\, \text{Tr}\{(\sigma^{j})^\dagger \rho_{ss}\} \, \text{Tr}\{\sigma^\dagger e^{\tau\mathcal{L}_{\text{LT}}} \left[\sigma^{j},\rho_{ss}\right]\}\\
s_{j}(\tau,\rho_{ss})=&\,\text{Tr}\{(\sigma^{j})^\dagger e^{\tau\mathcal{L}_{\text{LT}}}  \left[\sigma^\dagger,\rho_{ss}\right]\}\\
d_{j}(\tau,\rho_{ss})=&\,\text{Tr}\{(\sigma^{j})^\dagger  e^{\tau\mathcal{L}_{\text{LT}}} \sigma^\dagger \rho_{ss}\}\\
&\;\;\;\;\;\;\;\;\;\; - \text{Tr}\{(\sigma^{j})^\dagger  e^{\tau\mathcal{L}_{\text{LT}}} \rho_{ss}\} \, \text{Tr}\{\sigma^\dagger\rho_{ss}\}.
\end{split}
\end{equation}

\section{Notes on the superposition principle}\label{App superposition}

Since the Liouville equation, Eq.~(\ref{EOM main text}) of the main text, obeys a superposition principle, one might wonder whether this is still respected by the non-linear equations of the approach we derive.

In this context we first note that Eq.~(\ref{reduced NZE main text}) of the main text does no longer allow for a superposition principle for $R$, in contrast to Eq.~(\ref{EOM main text}) of the main text. Nonetheless, for two 
reduced density matrices $\rho_{n_{0}}$ and $\rho_{n_{0}}'$ that are solutions of Eq.~(\ref{reduced NZE main text}), their convex sum $c \rho_{n_{0}} + (1-c)\rho_{n_{0}}'$ ($0\le c\le 1$) is 
also a solution since Eq.~(\ref{reduced NZE main text}) 
is exact and its solutions are thus identical to $\text{Tr}_{\not{n_{0}}}R$. Despite its nonlinearity, Eq.~(\ref{reduced NZE main text}) thus fulfills a superposition principle for reduced density matrices $\rho_{n_{0}}$. 
Due to the applied approximations, the superposition principle of Eq.~(\ref{reduced NZE main text}) for $\rho_{n_{0}}$ does not necessarily hold for Eq.~(\ref{NZE in BA main}) of the main text. We have confirmed that it does hold 
for systems where we found Eq.~(\ref{NZE in BA main}) to become exact, e.g. examples of two coupled spins, but in general, the superposition principle is lost on time scales where Eq.~(\ref{NZE in BA main}) 
ceases to be a good approximation.

Moreover, the stationary states of Eq.~(\ref{EOM main text}) of the main text for the spin model [Eq.~(\ref{full BH dynamics}) of the main text] are expected to 
be unique \cite{Spohn,Schirmer} and one should not expect a 
superposition principle for Eq.~(\ref{eq:c-MoP-main-text}) of the main text.

\section{Bistability within single-site mean-field solutions}\label{App bistability}
For the stationary states we find a bistabilty in the single-site mean-field solution due to the non-linear character of the mean-field equation, which is no longer present in the 
two-site cluster version, see Fig. 3 of the main text.  
The c-MoP equation [Eq.~(\ref{eq:c-MoP-main-text}) of the main text] which is a non-linear algebraic equation as well, does not exhibit bistable behavior for any of the considered variants in the 
whole parameter 
range of our study. This is in agreement with Spohn's theorem \cite{Spohn,Schirmer} which suggests that
the dynamics of a Lindblad type equation of motion \cite{Lindblad}, just like Eq.~(\ref{EOM main text}) of the main text with the $\mathcal{L}_{n}$ and $\mathcal{L}_{I}$ as 
specified in Eq.~(\ref{full BH dynamics}), 
relaxes to a 
unique steady-state. Hence bistabilities resulting from mean-field calculations \cite{Boite12} can in general not be attributed a physical existence.


\begin{thebibliography}{99}
%
\bibitem{Sachdev11}
S. Sachdev,
{\em Quantum Phase Transitions},
Cambridge University Press (2011).
%
\bibitem{Hasan10}
M. Z. Hasan and C. L. Kane,
Rev. Mod. Phys. {\bf 82}, 3045 (2010).
%
\bibitem{Nayak08}
C. Nayak, S. H. Simon, A. Stern, M. Freedman, and S. Das Sarma,
Rev. Mod. Phys. {\bf 80}, 1083 (2008).
%
\bibitem{Leggett06}
A. J. Leggett,
Nature Phys. {\bf 2}, 134 (2006).
%
\bibitem{Schollwoeck05}
U. Schollw\"ock,
Rev. Mod. Phys. {\bf 77}, 259 (2005).
%
\bibitem{Kadanoff09}
L. P. Kadanoff,
J. Stat. Phys. {\bf 137}, 777 (2009).
%
\bibitem{Fisher89}
M. P. A. Fisher, P. B. Weichman, G. Grinstein, and D. S. Fisher,
Phys. Rev. B {\bf 40}, 546 (1989).
%
\bibitem{Mori65}
H. Mori,
Prog. Theor. Phys. {\bf 33}, 423 (1965).
%
\bibitem{Zwanzig01}
R. Zwanzig,
{\em Nonequilibrium Statistical Mechanics},
Oxford University Press, Oxford (2001).
%
\bibitem{Breuer07}
H.-P. Breuer and F. Petruccione,
{\em The Theory of Open Quantum Systems},
Oxford University Press (2007).
%
\bibitem{FZ01}
        R.~Fazio and H.S.J.~van~der~Zant,
        Phys.~Rep. {\bf 355}, 235 (2001).
%
\bibitem{BDZ07}
        I.~Bloch, J.~Dalibard and W.~Zwerger,
        Rev. Mod. Phys. {\bf 80}, 885 (2008).
%
\bibitem{ion}
R. Islam, C. Senko, W. C. Campbell, S. Korenblit, J. Smith, A. Lee, E. E. Edwards, C.-C. J. Wang, J. K. Freericks, and C. Monroe,
Sience {\bf 340}, 583 (2013).
%
\bibitem{Barreiro11}
J. T. Barreiro, M. M\"uller, P. Schindler,	 D. Nigg, T. Monz, M. Chwalla, M. Hennrich, C. F. Roos, P. Zoller, and R. Blatt,
Nature {\bf 470}, 486 (2011).
%
\bibitem{Hartmann06}
  M. J. Hartmann, F. G. S. L. Brand\~ao, and M. B. Plenio,
  Nature Phys. {\bf 2}, 849 (2006).
%
\bibitem{Kollath07}
C. Kollath, A. M. L\"auchli, and E. Altman,
Phys. Rev. Lett. {\bf 98}, 180601 (2007)
%
\bibitem{Trotzky12}
S. Trotzky, Y-A. Chen, A. Flesch, I. P. McCulloch, U. Schollw\"ock, J. Eisert, and I. Bloch,
Nature Phys. {\bf 8}, 325 (2012).

\bibitem{Prosen08}
T. Prosen and I. Pi\v{z}orn,
Phys. Rev. Lett. {\bf 101}, 105701 (2008).
%
\bibitem{Hartmann10}
M. J. Hartmann,
Phys. Rev. Lett., {\bf 104}, 113601 (2010).
%
\bibitem{Nissen12}
F. Nissen, S. Schmidt, M. Biondi, G. Blatter, H.E. T\"ureci, and J. Keeling,
Phys. Rev. Lett. {\bf 108}, 233603 (2012).
%
\bibitem{Diehl2008}
  S. Diehl, A. Micheli, A. Kantian, B. Kraus, H.P. B\"uchler, and P. Zoller, 
  Nature Phys. {\bf 4}, 878 (2008).
%
\bibitem{Diehl2010}
  S. Diehl, A. Tomadin, A. Micheli, R. Fazio, and P. Zoller,
  Phys. Rev. Lett. {\bf 105}, 015702 (2010).
%
\bibitem{Diehl11}
S. Diehl, E. Rico, M. A. Baranov, and P. Zoller,
Nature Phys. {\bf 7}, 971 (2011).
%
%
\bibitem{Rivas12}
A. Rivas and S. F. Huelga, 
{\em Open Quantum Systems. An Introduction}, Springer, (2011).
\bibitem{Lindblad}
G. Lindblad, 
Comm. Math. Phys. 48, {\bf 119} (1976).
%
%
\bibitem{Intdiff}
V. Lakshmikantham, M. Rama Mohana Rao, {\em Theory of Integro-Differential Equations}, CRC Press (1995).
%
\bibitem{Cirac92}
J. I. Cirac, R. Blatt, P. Zoller, and W. D. Phillips, Phys. Rev. A {\bf 46}, 2668 (1992)
%
\bibitem{Rokhsar91}
D. S. Rokhsar and B. G. Kotliar,
Phys. Rev. B {\bf 44}, 10328 (1991).
%
\bibitem{Jin13}
J. Jin, D. Rossini, R. Fazio, M. Leib, and M. J. Hartmann,
Phys. Rev. Lett. {\bf 110}, 163605 (2013).
%
\bibitem{Flesch08}
A. Flesch, M. Cramer, I. P. McCulloch, U. Schollw\"ock, and J. Eisert,
Phys. Rev. A {\bf 78}, 033608 (2008)
%
\bibitem{Hartmann09}
M. J. Hartmann, J. Prior, S. R. Clark and M.B. Plenio, 
Phys. Rev. Lett., {\bf 102}, 057202 (2009).
%
\bibitem{Nielsen}
M.A. Nielsen and I.L. Chuang, {\em Quantum Computation and Quantum Information}, Cambridge University Press 2000.
%
\bibitem{Spohn}
H. Spohn,
Lett. Math. Phys. 2, {\bf 33} (1977).
%
\bibitem{Schirmer}
S. G. Schirmer and Xiaoting Wang, 
Phys. Rev. A {\bf 81}, 062306 (2010).
%
\bibitem{delValle13}
E. del Valle and M.J. Hartmann, 
J. Phys. B: At. Mol. Opt. Phys. {\bf 46}, 224023 (2013).
%
\bibitem{Li13}
Andy C.Y. Li, F. Petruccione, and Jens Koch,
arXiv:1311.3227 (2013)
%
\bibitem{delValle12}
E. del Valle, A. Gonzalez-Tudela, F.P. Laussy, C. Tejedor and M.J. Hartmann, 
{\em Theory of frequency-filtered and time-resolved N-photon correlations},
Phys. Rev. Lett. {\bf 109}, 183601 (2012).
%
\bibitem{Boite12}
A. Le Boit\'{e}, G. Orso, and C. Ciuti,
Phys. Rev. Lett. {\bf 110}, 233601 (2013).
%
\end{thebibliography}
\end{document}